\documentclass[aps,prd,letterpaper,floats,floatfix,nofootinbib]{revtex4-1}
\usepackage{geometry,graphicx,amsmath,amssymb,amsfonts,hyperref}

\begin{document}

\title{Constraining models with a large scalar multiplet}

\author{Kevin Earl}
\email{kevinearl@cmail.carleton.ca}
\author{Katy Hartling\footnote{Formerly Katy Hally.}}
\email{khally@physics.carleton.ca}
\author{Heather E.~Logan}
\email{logan@physics.carleton.ca}
\author{Terry Pilkington}
\email{tpilking@physics.carleton.ca}

\affiliation{Ottawa-Carleton Institute for Physics, Carleton University, Ottawa, Ontario K1S 5B6, Canada}

\date{March 5, 2013}

\begin{abstract}
Models in which the Higgs sector is extended by a single electroweak scalar multiplet $X$ can possess an accidental global U(1) symmetry at the renormalizable level if $X$ has isospin $T \geq 2$.  We show that all such models with an accidental U(1) symmetry are excluded by the interplay of the cosmological relic density of the lightest (neutral) component of $X$ and its direct-detection cross section via $Z$ exchange.  The sole exception is the $T=2$ multiplet, whose lightest member decays on a few-day to few-year time scale via a Planck-suppressed dimension-5 operator.
\end{abstract}

\maketitle
\section{Introduction}

Extensions of the scalar sector of the Standard Model (SM) beyond the minimal single Higgs doublet are of great interest in model building and collider phenomenology and are, as yet, largely unconstrained by experiment.
Such extensions are common in models that address the hierarchy problem of the SM, such as supersymmetric models~\cite{SUSY} and little Higgs models~\cite{LH}, as well as in models for neutrino masses, dark matter, etc.  Most of these extensions contain additional SU(2)$_L$-singlet, -doublet, and/or -triplet scalar fields. However, some extensions of the SM contain scalars in larger multiplets of SU(2)$_L$.  Such larger multiplets have been used to produce a natural dark matter candidate~\cite{Cirelli:2005uq,Cirelli:2009uv,Cai:2012kt}, which is kept stable thanks to an accidental global symmetry present in the Higgs potential for multiplets with isospin $T \geq 2$.  Three different models with a Higgs quadruplet ($T=3/2$) have also been proposed for neutrino mass generation~\cite{Babu:2009aq,Picek:2009is,Ren:2011mh}.  Models in which the SM SU(2)-doublet Higgs mixes with a seven-plet ($T = 3$), aided by additional representations of SU(2), have also been studied recently in Ref.~\cite{Hisano:2013sn}.

In this paper we consider models that extend the SM scalar sector through the addition of a \emph{single} large multiplet.  For multiplets with $n \equiv 2T+1 \geq 5$ (isospin 2 and larger), the scalar potential of these models always preserves an accidental global U(1) or $Z_2$ symmetry at the renormalizable level.  If unbroken, such a symmetry forces the lightest member of the large multiplet to be stable.  Spontaneous breaking of an accidental global U(1) symmetry is phenomenologically unacceptable because it would lead to a massless Goldstone boson that couples to fermions through its mixing with the SM Higgs doublet's neutral Goldstone, and thus mediate new long-range forces between SM fermions.  Furthermore, perturbative unitarity of scattering amplitudes involving pairs of scalars and pairs of SU(2) gauge bosons requires that $T \leq 7/2$ (i.e., $n \leq 8$) for a complex scalar multiplet and $T \leq 4$ (i.e., $n \leq 9$) for a real scalar multiplet~\cite{Hally:2012pu}.

The models that preserve such an accidental global symmetry can be grouped into three classes based on the hypercharge $Y$ of the large multiplet, as follows:
\begin{enumerate}
\item[$(i)$] Models with a $Y=0$ multiplet, with $n = 5$, 7, or 9, corresponding to isospin 2, 3, or 4.\footnote{Note that a real multiplet must have integer isospin, and a complex $Y=0$ multiplet can always be written in terms of two real $Y=0$ multiplets.}  In the most general case the large multiplet is odd under an accidental global $Z_2$ symmetry; though an additional U(1) symmetry may be imposed by hand~\cite{Joachim}, it will not be accidental. These models have previously been considered in Refs.~\cite{Cirelli:2005uq,Cirelli:2009uv} as possible candidates for ``next-to-minimal'' dark matter. 
\item[$(ii)$] Models with a complex multiplet with $n = 5$, 6, 7, or 8, with $Y = 2T$ (we work in the convention $Q = T^3 + Y/2$).  The large multiplet is charged under an accidental global U(1) symmetry.  The hypercharge is chosen so that the lightest member of the multiplet can be electrically neutral.\footnote{Models in which the lightest member of the large multiplet is electrically charged are excluded or strongly constrained by the absence of electrically charged relics.  Metastable multicharged states are constrained by direct collider searches to be heavier than about 400--500~GeV, depending on their charge~\cite{ATLAS:charged}.}  The masses of the states in the large multiplet are split by an operator of the form $(\Phi^{\dagger} \tau^a \Phi) (X^{\dagger} T^a X)$, where $\Phi$ is the SM Higgs doublet, $X$ is the large multiplet, and $\tau^a$ and $T^a$ are the appropriate SU(2) generators.  We study these models in the current paper.
\item[$(iii)$] Models with a complex multiplet with $n = 6$ or 8, with $Y = 1$.  The large multiplet is odd under an accidental global $Z_2$ symmetry.  The would-be accidental global U(1) symmetry is broken by an operator of the form $(\Phi^{\dagger} \tau^a \widetilde \Phi)(\widetilde X^{\dagger} T^a X)$, where $\widetilde \Phi$, $\widetilde X$ denote the conjugate multiplets.  Such an operator can appear only when $n$ is even.  We will address these models in a forthcoming paper~\cite{Z2}.
\end{enumerate}

In this paper we study the constraints on the models with a complex multiplet with $n = 5$, 6, 7, or 8 and $Y = 2T$. We first determine the constraints on the scalar quartic couplings from perturbative unitarity and precision electroweak measurements.  We then examine the bounds on the neutral scalar $\chi^0$ from cosmological considerations.  Our goal is not to determine whether $\chi^0$ is able to account for the entire observed quantity of dark matter---this possibility is strongly excluded by dark matter direct-detection experiments---but rather to evaluate the ultimate viability of the model as a target for collider searches.  Assuming a standard thermal history of the universe, we compute the thermal relic density of $\chi^0$ and $\chi^{0*}$ and compare it with the limits from dark matter direct-detection experiments.  In conjunction with the requirement that $m_{\chi^0} \gtrsim m_Z/2 \simeq 45~{\rm GeV}$ from the invisible width of the $Z$ boson, we find that these cosmological constraints exclude the models with $n = 6$, 7, and 8.

The $n=5$ ($T=2$) multiplet has a dimension-5 Planck-suppressed interaction with the SM Higgs doublet of the form 
\begin{equation}
	\mathcal{L} \supset \frac{1}{M_{Pl}} \Phi \Phi \Phi \Phi X^{\dagger} + {\rm h.c.},
\end{equation}
where $M_{Pl}$ is the Planck mass.  This operator induces a mixing of the SM Higgs into the neutral component of $X$, $\chi^{\prime} = \chi^{0,r} - \epsilon \phi^{0,r}$, with $\epsilon \sim v^3 / [(m_{\chi^0}^2 - m_h^2) M_{Pl}]$.  Here $v \simeq 246$~GeV is the SM Higgs vacuum expectation value.  The neutral member of $X$ can then decay via its Higgs component.  Assuming that $\chi^0$ is the lightest state, its lifetime ranges from a few days to a few years for $m_{\chi^0} \sim 100$--1000~GeV.  This puts the decays of the lightest neutral state of the $n=5$ model well after big-bang nucleosynthesis and well before the recombination surface of the cosmic microwave background radiation.  Direct detection constraints from present-day experiments therefore do not apply to this model, and it remains viable~\cite{Joachim}.

The paper is organized as follows.  In Sec.~\ref{sec:models} we give the Lagrangians and mass spectra for the four models that we consider.  In Sec.~\ref{sec:constraints} we obtain the indirect constraints on the model parameters from unitarity and the oblique parameters, and comment on collider constraints.  In Sec.~\ref{sec:dirdet} we calculate the upper bound on the relic density of neutral scalars $\chi^0$, $\chi^{0*}$ from dark matter direct-detection experiments.  In Sec.~\ref{sec:relic} we compute the relic density from thermal freeze-out and show that in all cases it yields a relic density too large to be consistent with the bound from direct detection.
We conclude in Sec.~\ref{sec:conclusions}.  Feynman rules, formulas for the oblique parameters, and formulas for the partial decay width of the SM Higgs to two photons are collected in the appendices.

\section{The models}
\label{sec:models}

The models that we consider extend the SM through the addition of a single complex scalar multiplet $X$, with hypercharge $Y = 2T$ (normalized so that $Q = T^3 + Y/2$).  The hypercharge is chosen so that the lightest member of $X$ can be neutral. 
The size of the multiplet $X$ is restricted to $n \equiv 2T + 1 \leq 8$ by the requirement that tree-level amplitudes for SU(2) gauge bosons scattering into the states in $X$, $W^a W^a \to \chi^{Q*} \chi^Q$, remain perturbative~\cite{Hally:2012pu}.  When $n \geq 5$, the scalar potential possesses an accidental global U(1) symmetry corresponding to phase rotations of $X$.  This U(1) symmetry ensures that the lightest member of $X$ is stable, at least at the level of renormalizable operators.  These two conditions leave us with four models to consider, with $n = 5$, 6, 7, and 8.  

The gauge-invariant scalar potential is given by
\begin{equation}
	V(\Phi,X) = m^2 \Phi^\dagger \Phi + M^2 X^\dagger X 
	+ \lambda_1 \left(\Phi^\dagger \Phi \right)^2 
	+ \lambda_2 \Phi^\dagger \Phi X^\dagger X 
	+ \lambda_3 \Phi^\dagger \tau^a \Phi X^\dagger T^a X + \mathcal{O}(X^4),
	\label{eq:complexpotential}
\end{equation}
where $\tau^a$ and $T^a$ are the generators of SU(2)$_L$ in the doublet and $n$-plet representations, respectively, $\Phi$ is the usual SM Higgs doublet, and the large scalar multiplet takes the form 
\begin{equation}
	X = (\chi^{+(n-1)},\cdots,\chi^0)^T.  
\end{equation}	

The mass of particle $\chi^Q$ with charge $Q = T^3 + Y/2 \geq 0$ is given by
\begin{equation}
	m_{\chi^Q}^2 = M^2 + \frac{1}{2} v^2 \left[ \lambda_2 - \frac{1}{2} \lambda_3 T^3 \right]
	= M^2 + \frac{1}{2}v^2
	\left[\lambda_2 - \frac{1}{2}\lambda_3\left(Q - \frac{n - 1}{2}\right)\right]
	\equiv M^2 + \frac{1}{2} v^2 \Lambda_Q,
\end{equation}
where $v \simeq 246$~GeV is the SM Higgs vacuum expectation value (vev)
and we define the dimensionless couplings $\Lambda_Q$ as the quantity in square brackets above.
The neutral particle $\chi^0$, with $T^3 = -T = -(n-1)/2$, will have a mass,
\begin{equation}
	m_{\chi^0}^2 = M^2 + \frac{1}{2}v^2\left[\lambda_2 + \frac{1}{4}\lambda_3 (n - 1) \right] 
	= M^2 + \frac{1}{2}v^2\Lambda_0.
	\label{eq:mass}
\end{equation}
The masses of the charged states $\chi^Q$ can be written in terms of the $\chi^0$ mass as
\begin{equation}
	m_{\chi^Q}^2 = m_{\chi^0}^2 -\frac{1}{4} v^2 \lambda_3 Q.
\end{equation}
We require that the stable lightest member of $X$ is electrically neutral; this forces us to take $\lambda_3 < 0$.

\section{Constraints on couplings and masses}
\label{sec:constraints}

\subsection{Unitarity constraints on scalar quartic couplings}

The scalar quartic couplings $\lambda_2$ and $\lambda_3$ given in Eq.~(\ref{eq:complexpotential}) can be bounded by requiring perturbative unitarity of the zeroth partial wave amplitude.  The partial wave amplitudes are related to scattering matrix elements according to
\begin{equation}
	\mathcal{M} = 16\pi \sum_J (2J+1) a_J P_J(\cos\theta),
\end{equation}
where $J$ is the orbital angular momentum of the final state and $P_J(\cos\theta)$ is the corresponding Legendre polynomial.
Perturbative unitarity of the zeroth partial wave amplitude dictates the tree-level constraint,
\begin{equation}
	|{\rm Re} \, a_0 | \leq \frac{1}{2}.
	\label{rea0}
\end{equation}

The coupling $\lambda_2$ controls the strength of the isospin-zero $\chi^* \chi \to \phi^* \phi$ amplitude, while $\lambda_3$ controls the strength of the isospin-one $\chi^* \chi \to \phi^* \phi$ channel.  Because we are working with large scalar multiplets, the isospin-zero $\chi^* \chi \to WW,BB$ and isospin-one $\chi^* \chi \to WB$ amplitudes can be significant~\cite{Hally:2012pu}, leading to more stringent coupled-channel limits on $\lambda_2$ and $\lambda_3$.  We neglect all other contributing processes\footnote{Additional contributions to the matrix of coupled-channel amplitudes come from quartic couplings of $X$ as well as $\phi^*\phi \to \phi^*\phi$ amplitudes proportional to $\lambda_1$.  We find numerically that including these contributions generically leads to a slightly tighter constraint on $\lambda_2$ and $\lambda_3$, but this constraint depends on the interplay between the $\lambda_1$ contributions and those from the quartic $X$ couplings.} and work in the high-energy limit.

The relevant amplitudes for the isospin-zero channels are
\begin{eqnarray}
	a_0([\chi^*\chi]_0 \to [\phi^*\phi]_0) &=& - \frac{\sqrt{n}}{8 \sqrt{2} \pi}\lambda_2, \nonumber \\
	a_0([\chi^*\chi]_0 \to [WW]_0) &=& \frac{g^2}{16 \pi} \frac{(n^2-1) \sqrt{n}}{2 \sqrt{3}}, \nonumber \\
	a_0([\chi^*\chi]_0 \to [BB]_0) &=& \frac{g^2}{16 \pi} \frac{s^2_W}{c^2_W} \frac{Y^2 \sqrt{n}}{2} 
	= \frac{g^2}{16 \pi} \frac{s^2_W}{c^2_W} \frac{(n-1)^2 \sqrt{n}}{2},
\end{eqnarray}
where the $\chi^*\chi \to WW,BB$ amplitudes include both of the contributing transverse gauge boson polarization combinations~\cite{Hally:2012pu} and we used $Y = 2T = n-1$ in the last line.  Here $g$ is the SU(2)$_L$ gauge coupling and $s_W, c_W \equiv \sin\theta_W, \cos\theta_W$ are the sine and cosine of the weak mixing angle.  We define the following normalized isospin-zero field combinations,
\begin{eqnarray}
	[\phi^*\phi]_0 &=& \frac{1}{\sqrt{2}}(\phi^+\phi^- + \phi^{0*}\phi^0), \nonumber \\ {}
	[\chi^*\chi]_0 &=& \frac{1}{\sqrt{n}}\sum_{Q=0}^{n-1} \chi^{Q*} \chi^Q, \nonumber \\ {}
	[WW]_0 &=& \frac{1}{\sqrt{3}}\left(\sqrt{2} W^+W^- + \left(\frac{W^3 W^3}{\sqrt{2}}\right)\right),
		\nonumber \\ {}
	[BB]_0 &=& BB/\sqrt{2}.
\end{eqnarray}
The relevant amplitudes for the isospin-one channels are
\begin{eqnarray}
	a_0([\chi^*\chi]_1 \to [\phi^*\phi]_1) &=& - \frac{\sqrt{n(n^2-1)}}{32 \sqrt{6} \pi} \lambda_3, \nonumber \\
	a_0([\chi^*\chi]_1 \to [WB]_1) 
	&=& \frac{g^2}{16 \pi} \frac{s_W}{c_W} \frac{Y \sqrt{n (n^2 - 1)}}{\sqrt{6}} 
	= \frac{g^2}{16 \pi} \frac{s_W}{c_W} \frac{(n-1) \sqrt{n (n^2 - 1)}}{\sqrt{6}},
\end{eqnarray}
where again the $\chi^*\chi \to WB$ amplitude includes both of the contributing transverse gauge boson polarization combinations~\cite{Hally:2012pu}.  Here we used the following normalized isospin-one field combinations,
\begin{eqnarray}
	[\phi^*\phi]_1 &=& \frac{1}{\sqrt{2}}(\phi^+\phi^- - \phi^{0*}\phi^0), \nonumber \\ {}
	[\chi^*\chi]_1 &=& \sqrt{\frac{12}{n(n^2-1)}} \sum_{Q=0}^{n-1} \chi^{Q*} T^3 \chi^Q
		= \sqrt{\frac{12}{n(n^2-1)}} \sum_{Q=0}^{n-1} \chi^{Q*} \left[Q - \frac{n-1}{2} \right] \chi^Q,  \nonumber \\ {}
	[WB]_1 &=& W^3 B.
\end{eqnarray}

Finding the largest eigenvalue of each coupled-channel matrix and applying the unitarity constraint of Eq.~(\ref{rea0}), we find the unitarity bounds on $\lambda_2$ and $\lambda_3$,
\begin{eqnarray}
	|\lambda_2| &\leq& \sqrt{\frac{32 \pi^2}{n} - g^4\frac{(n^2-1)^2}{24} - g^4 \frac{s_W^4}{c_W^4} \frac{(n-1)^4}{8}}, 
	\nonumber \\
	|\lambda_3| &\leq& 2\sqrt{\frac{384\pi^2}{{n(n^2-1)}} - g^4 \frac{s_W^2}{c_W^2} (n-1)^2}.
\end{eqnarray}
Recall that $\lambda_3$ must be negative so that $\chi^0$ is the lightest member of the large multiplet.  $\lambda_2$ can have either sign.
Numerical bounds are given for $n = 5$, 6, 7, and 8 in Table~\ref{lambdas}.\footnote{In this table we use $g^2= 4\pi \alpha/ s_W^2$, $s_W^2=0.231$, and $\alpha = 1/128$.}

\begin{table}
\begin{tabular}{ccc}
\hline \hline
\ $n$ \ & \ $|\lambda_2^{\rm lim}|$ \ & \ $|\lambda_3^{\rm lim}|$ \ \\
\hline
5 & 7.64 & 11.1  \\
6 & 6.49 & 8.17  \\
7 & 5.01 & 6.11  \\
8 & 2.17 & 4.41  \\
\hline \hline
\end{tabular}
\caption{Upper limits on $|\lambda_2|$ and $|\lambda_3|$ from perturbative unitarity, for $Y = 2T = n-1$.}
\label{lambdas}
\end{table}

\subsection{Electroweak precision constraints}

The multiplet $X$ contributes to electroweak observables through the oblique parameters $S$, $T$, and $U$~\cite{Peskin:1990zt}.  The contributions of such a scalar multiplet obeying a U(1) global symmetry, so that the mass eigenstates have definite $T^3$, were computed for arbitrary isospin and hypercharge in Ref.~\cite{Lavoura:1993nq}.  We summarize the results in Appendix~\ref{sec:oblique} for completeness.  

The contributions to the oblique parameters depend only on $n$, $Y$, and the masses $m_{\chi^Q}$ of each state in the multiplet.  Therefore, for a given $n$ and setting $Y = 2T = n-1$, the oblique parameters constrain only two model parameters, which can be chosen as $m_{\chi^0}$ (which sets the overall mass scale) and $\lambda_3$ (which controls the mass splittings).

\begin{figure}
\resizebox{0.5\textwidth}{!}{
\includegraphics{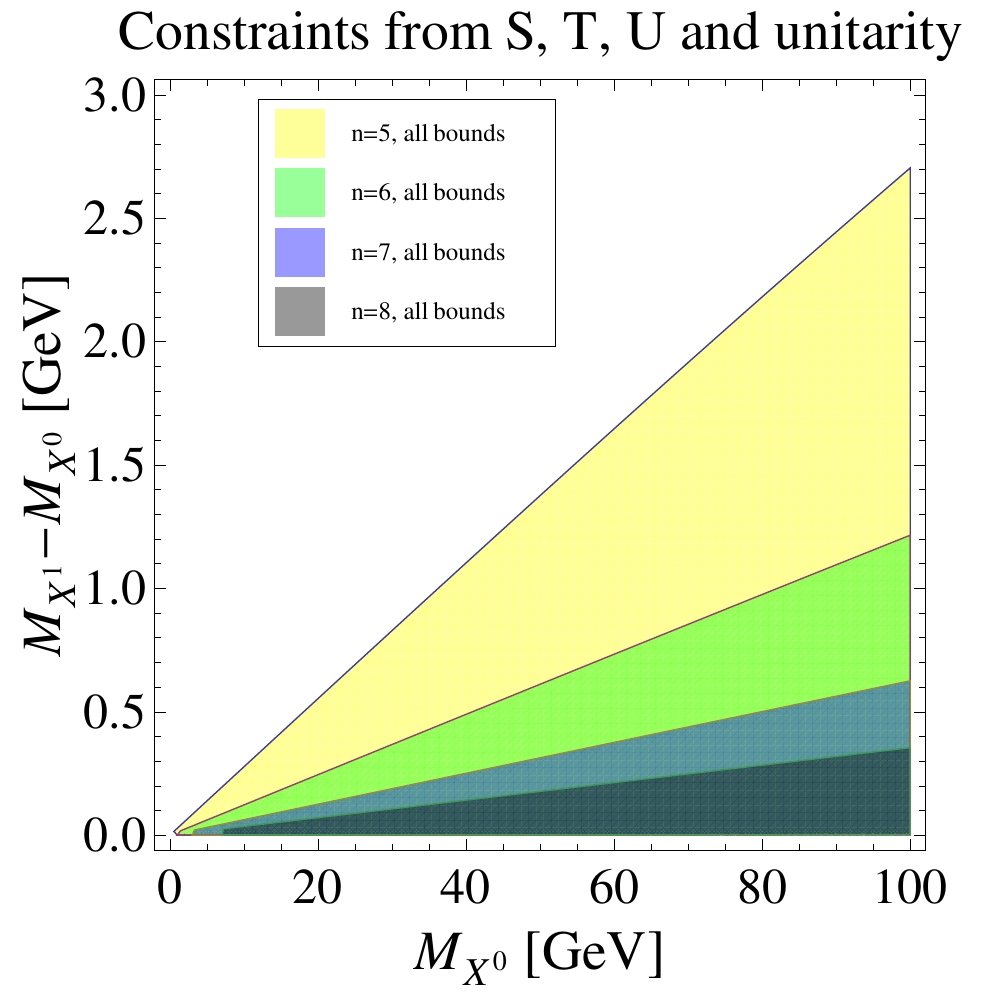}
}\resizebox{0.5\textwidth}{!}{
\includegraphics{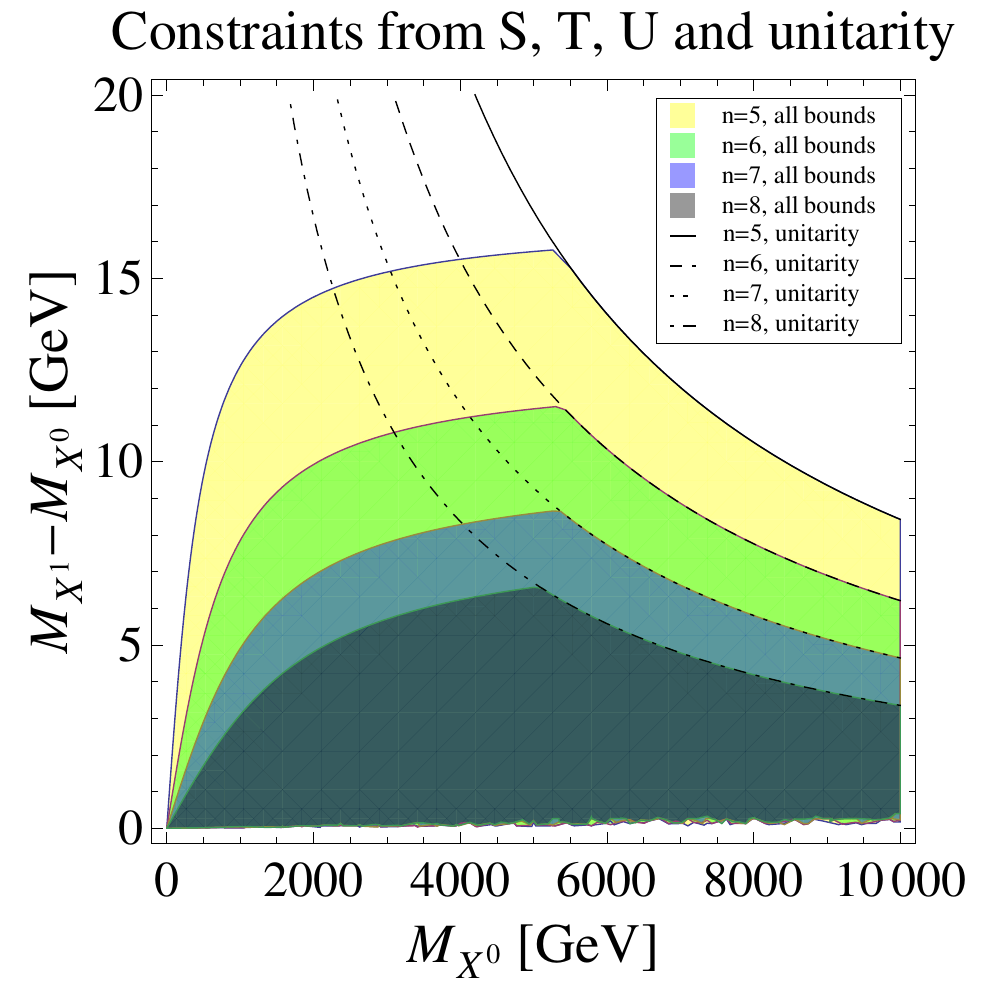}}
\caption{The 95\% confidence level constraints on $\Delta m \equiv m_{\chi^{+1}} - m_{\chi^0}$ as a function of $m_{\chi^0}$ from the $S$, $T$, and $U$ parameters, for the scalar multiplets with $n = 5$, 6, 7, and 8 and $Y = 2T = n - 1$.  Dashed lines indicate the upper limit on $\Delta m$ from the unitarity bound on $\lambda_3$.  The left panel shows the low-$m_{\chi^0}$ region while the right panel extends to higher masses.
\label{XSTU}}
\end{figure}

The current experimental values relative to the SM with Higgs mass $m_h = 126$~GeV 
are $S_{\rm exp}=0.03\pm0.10$, $T_{\rm exp}=0.05\pm0.12$, $U_{\rm exp}=0.03\pm0.10$, with relative correlations of $\rho_{ST}=0.89$, $\rho_{TU}=-0.83$, $\rho_{SU}=-0.54$~\cite{STU2012}.  We use these values to constrain $m_{\chi^0}$ and $\lambda_3$ via a two-parameter $\chi^2$ variable; for details see Appendix~\ref{sec:oblique}. We show 95\% confidence level ($\chi^2 = 5.99$) limits for the two parameters $m_{\chi^0}$ and $\Delta m \equiv m_{\chi^+} - m_{\chi^0}$ in Fig.~\ref{XSTU}.  

At low $m_{\chi^0}$ the constraint is dominated by the $S$ parameter and leads to an upper bound on $\Delta m$ that is linear in $m_{\chi^0}$.  This bound can be parametrized as 
\begin{equation}
	\Delta m \equiv m_{\chi^+} - m_{\chi^0}
	= 0.031 \left[ \frac{1}{n-4} - 0.13 \right] m_{\chi^0}.
	\label{eq:STUparam}
\end{equation}
For larger $m_{\chi^0} \sim 1$~TeV, the constraint from the $T$ parameter becomes important and limits the value of $\Delta m$ independent of $m_{\chi^0}$.  For $m_{\chi^0} \sim 5$--6~TeV, the unitarity limit on $\lambda_3$ becomes the strongest constraint on $\Delta m$, as shown by the dashed lines in Fig.~\ref{XSTU}.  

Notice that for $n \geq 6$ the mass splitting between $\chi^0$ and the next-lightest state $\chi^+$ is constrained to be no more than about 12~GeV; for $m_{\chi^0} \sim 100$~GeV the splitting is less than 1.5~GeV.  The maximum allowed mass splitting decreases with increasing $n$.

\subsection{Direct collider constraints}

Scalar particle masses below about 100~GeV are constrained by $\chi^Q \chi^{Q*}$ pair production in $e^+e^-$ collisions at the CERN Large Electron-Positron (LEP-II) collider.  However, a dedicated search for the decay signatures in the models we consider here has not been made; furthermore, $\chi^Q \chi^{Q*}$ events may be difficult to detect if the mass splittings are small, leading to low-energy charged particles from the $\chi^Q$ decays.

Regardless of these difficulties, the LEP-I measurement of the $Z$ boson invisible decay width puts a stringent constraint on $Z \to \chi^0 \chi^{0*}$ independent of the mass splittings.  This leads to the requirement 
\begin{equation}
	m_{\chi^0} \gtrsim m_Z/2 \simeq 45~{\rm GeV}.  
\end{equation}	

Scalar particle masses below $m_h/2 \simeq 63$~GeV are also constrained by measurements of Higgs production and decay at the CERN Large Hadron Collider (LHC).  The decay width of the Higgs to $\chi^{Q*} \chi^Q$ is given by
\begin{equation}
	\Gamma(h \to \chi^{Q*} \chi^Q) = \frac{v^2 \Lambda_Q^2}{16 \pi m_h} 
	\sqrt{1 - \frac{4 m_{\chi^Q}^2}{m_h^2}}, \qquad {\rm for} \ m_{\chi^Q} \leq m_h/2,
\end{equation}
where $\Lambda_Q$ controls the $h \chi^{Q*} \chi^Q$ coupling (see Eq.~(\ref{eq:xxh})).

The ATLAS experiment has recently performed a direct search for $pp \to Zh$ with $h \to$~invisible, which found a 95\% confidence level upper bound ${\rm BR}(h \to {\rm invisible}) \leq 0.65$~\cite{ATLASinvh}, assuming the SM production rate for $pp \to Zh$.  This can be used to constrain $\Lambda_0$ from invisible $h \to \chi^{0*} \chi^0$, as long as other decays $h \to \chi^{Q*}\chi^Q$ (with $Q \neq 0$) can be neglected.  However, the electroweak precision observables tightly constrain the mass splittings among the states $\chi^Q$, so that Higgs decays to multiple $\chi^Q$ species are generally kinematically accessible for $\chi^0$ masses of more than a few GeV below half the Higgs mass.

Instead we take advantage of the fact that the measured Higgs signal strengths in a variety of channels at the LHC are in rough agreement with the SM predictions.  Because Higgs production in our model is SM-like, this constrains the allowable decay width of the Higgs to non-SM final states.  In particular, we have
\begin{equation}
	{\rm BR}(h \to {\rm new}) 
	= \frac{\Gamma_{\rm new}}{\Gamma_{\rm tot}^{SM} + \Gamma_{\rm new}}
	= 1 - \mu_i,
\end{equation}
where $\mu_i$ is the Higgs signal strength in any SM channel for which the Higgs decay width is the same as in the SM and $\Gamma_{\rm tot}^{SM} \simeq 4.1$~MeV for $m_h = 125$~GeV~\cite{Dittmaier:2012vm}.  Setting aside $h \to \gamma\gamma$, which can be modified by scalars $\chi^Q$ running in the loop (this will be addressed in Sec.~\ref{sec:higgspole}), we take as a rough lower bound $\mu_i \gtrsim 0.35$ from Higgs decays to $WW$ and $ZZ$~\cite{LHCcouplings} (a full fit of Higgs signal strengths in our model is beyond the scope of this paper).

We scan over $\lambda_2$, $\lambda_3$, and $m_{\chi^0}$ for each of the models, imposing the precision electroweak constraints on $\lambda_3$ as a function of $m_{\chi^0}$, and compute $\Gamma_{\rm new}$ from all kinematically accessible final states.  We find that the Higgs signal strength constrains $|\lambda_2| \lesssim 0.015$ for $m_{\chi^0} \leq 55$~GeV, with very little dependence on $n$.  Closer to threshold, the bound is somewhat loosened due to kinematic suppression of the new decays; for $m_{\chi^0} = 60$~GeV, we find $|\lambda_2| \lesssim 0.05$ for $n = 5$, with a stronger constraint for higher $n$ values due to the smaller mass splittings among the $\chi^Q$ states required by the precision electroweak constraints.

As we will show, the constraint $m_{\chi^0} \gtrsim m_Z/2$ together with cosmological considerations will be sufficient to exclude the models with $n = 6$, 7, and 8.
Further investigation of the direct collider constraints on the $n = 5$ model is beyond the scope of this paper.

\section{Dark matter direct detection constraint}
\label{sec:dirdet}

The allowable relic density of $\chi^{0(*)}$ is constrained by its non-observation in direct dark matter detection experiments.  In particular, the fraction of the ambient dark matter density that can be attributed to $\chi$ is bounded from above according to
\begin{equation}
	\frac{\Omega_{\chi}}{\Omega_{\rm DM}} 
	\leq \frac{\sigma_{\rm SI}^{\rm limit}}{\sigma_{\rm SI}^{\chi}},
\end{equation}
where $\sigma_{\rm SI}^{\chi}$ is the spin-independent, per-nucleon scattering cross section for $\chi^0$ or $\chi^{0*}$ and $\sigma_{\rm SI}^{\rm limit}$ is the experimental upper limit on the spin-independent dark matter scattering cross section obtained assuming the canonical ambient dark matter density.  The strongest experimental upper limit currently comes from the XENON100 experiment~\cite{Aprile:2012nq}.  We note that this limit will not apply to the model with $n = 5$ because in this model the relic $\chi^0$ particles decay away on a time scale short compared to the age of the universe.

Because $\chi^0$ is a complex scalar, it scatters off nucleons via both $Z$ and Higgs exchange.  The $Z$-exchange diagram yields a large scattering cross section, leading to very stringent constraints on $\Omega_{\chi}/\Omega_{\rm DM}$.  We compute the cross sections in the zero-velocity limit assuming equal densities of $\chi^0$ and $\chi^{0*}$, such that $\Omega_{\chi} \equiv \Omega_{\chi^0} + \Omega_{\chi^{0*}}$.  We have,
\begin{equation}
	\sigma_{\rm SI}^{\chi^0} = \sigma_Z^{\chi^0} + \sigma_h^{\chi^0} + \sigma_{\rm int}^{\chi^0},
\end{equation}
where $\sigma_{\rm int}^{\chi^0}$ is the interference between the $Z$- and Higgs-exchange diagrams.  For $\chi^{0*}$ we have $\sigma_{Z,h}^{\chi^{0*}} = \sigma_{Z,h}^{\chi^0}$ and $\sigma_{\rm int}^{\chi^{0*}} = - \sigma_{\rm int}^{\chi^0}$, so the interference term cancels in the total scattering rate for equal densities of $\chi^0$ and $\chi^{0*}$.  

The $Z$-exchange cross section for scattering off a single nucleon $N = p, n$ is given by 
\begin{equation}
	\sigma_Z^{\chi} = \frac{(f^V_N)^2 (n - 1)^2 m_{\chi^0}^2}{\pi v^2 m_Z^2}
	\frac{m_N^2}{(m_N + m_{\chi^0})^2},
	\label{eq:ddz}
\end{equation}
where $v \simeq 246$~GeV is the SM Higgs vev, $m_N$ is the nucleon mass, and the vector couplings of the $Z$ to the nucleon are given by the sum of the corresponding valence quark couplings,\footnote{These nucleon vector couplings do not receive any QCD corrections in the limit of zero momentum transfer due to the conservation of the vector current~\cite{CVC}.}
\begin{equation}
	f_p^V = \frac{2m_Z}{v}\left(\frac{1}{4} - s_W^2\right), \qquad
	f_n^V = \frac{2m_Z}{v}\left(- \frac{1}{4} \right).
\end{equation}
The axial-vector couplings do not contribute in the zero-velocity limit.  Notice that $\sigma_Z^{\chi}$ becomes independent of $m_{\chi^0}$ in the large-$m_{\chi^0}$ limit, where it is of order $m_N^2/v^4$.  Note also that $\sigma_Z^{\chi}$ is fixed with no free parameters once $m_{\chi^0}$ and the size of the multiplet $n$ are specified.

The $h$-exchange cross section for scattering off a single nucleon $N = p, n$ is given by
\begin{equation}
	\sigma_h^{\chi} = \frac{(f^h_N)^2 \Lambda_0^2 v^2}{4 \pi m_h^4}
	\frac{m_N^2}{(m_N + m_{\chi^0})^2},
	\label{eq:ddh}
\end{equation}
where $\Lambda_0 v$ is the $h\chi^0\chi^{0*}$ coupling defined in Eq.~(\ref{eq:mass}) and the Higgs-nucleon Yukawa couplings are given by~\cite{Ellis:2000ds}
\begin{equation}
	f_p^h = \frac{m_p}{v}(0.350 \pm 0.048), \qquad 
	f_n^h = \frac{m_n}{v}(0.353 \pm 0.049).
\end{equation}
Notice that $\sigma_h^{\chi}$ goes like $1/m_{\chi^0}^2$ in the large-$m_{\chi^0}$ limit, where it is of order $m_N^4/v^4 m_{\chi^0}^2$.  The Higgs-exchange contribution is thus generically much smaller than the $Z$-exchange contribution.  The Higgs-exchange contribution also depends on the parameter $\Lambda_0 = \lambda_2 + \lambda_3(n-1)/4$.  We obtain the least stringent upper bound on $\Omega_{\chi}/\Omega_{\rm DM}$ when $\Lambda_0 = 0$.

Because dark matter particles moving in the galactic halo have de Broglie wavelengths that are large compared to the size of a nucleus, the amplitudes for scattering off each nucleon add coherently.  This can be accounted for by replacing $f_N^V$ and $f_N^h$ in Eqs.~(\ref{eq:ddz}) and (\ref{eq:ddh}) above by the coherent nucleon-averaged values,
\begin{eqnarray}
	(f_N^V)^2 &\to& (\overline{f_N^V})^2 = \frac{[Z f_p^V + (A-Z) f_n^V]^2}{A^2}, \nonumber \\
	(f_N^h)^2 &\to& (\overline{f_N^h})^2 = \frac{[Z f_p^h + (A-Z) f_n^h]^2}{A^2} 
		\simeq (f_p^h)^2 \simeq (f_n^h)^2,
\end{eqnarray}
where $Z$ is the atomic number and $A$ the atomic mass of the nucleus.  For xenon, $Z = 54$ and $A$ ranges from 124 to 136.  We make a weighted average over the natural abundances of xenon isotopes~\cite{Krane}.

\begin{figure}
\resizebox{0.75\textwidth}{!}{\includegraphics{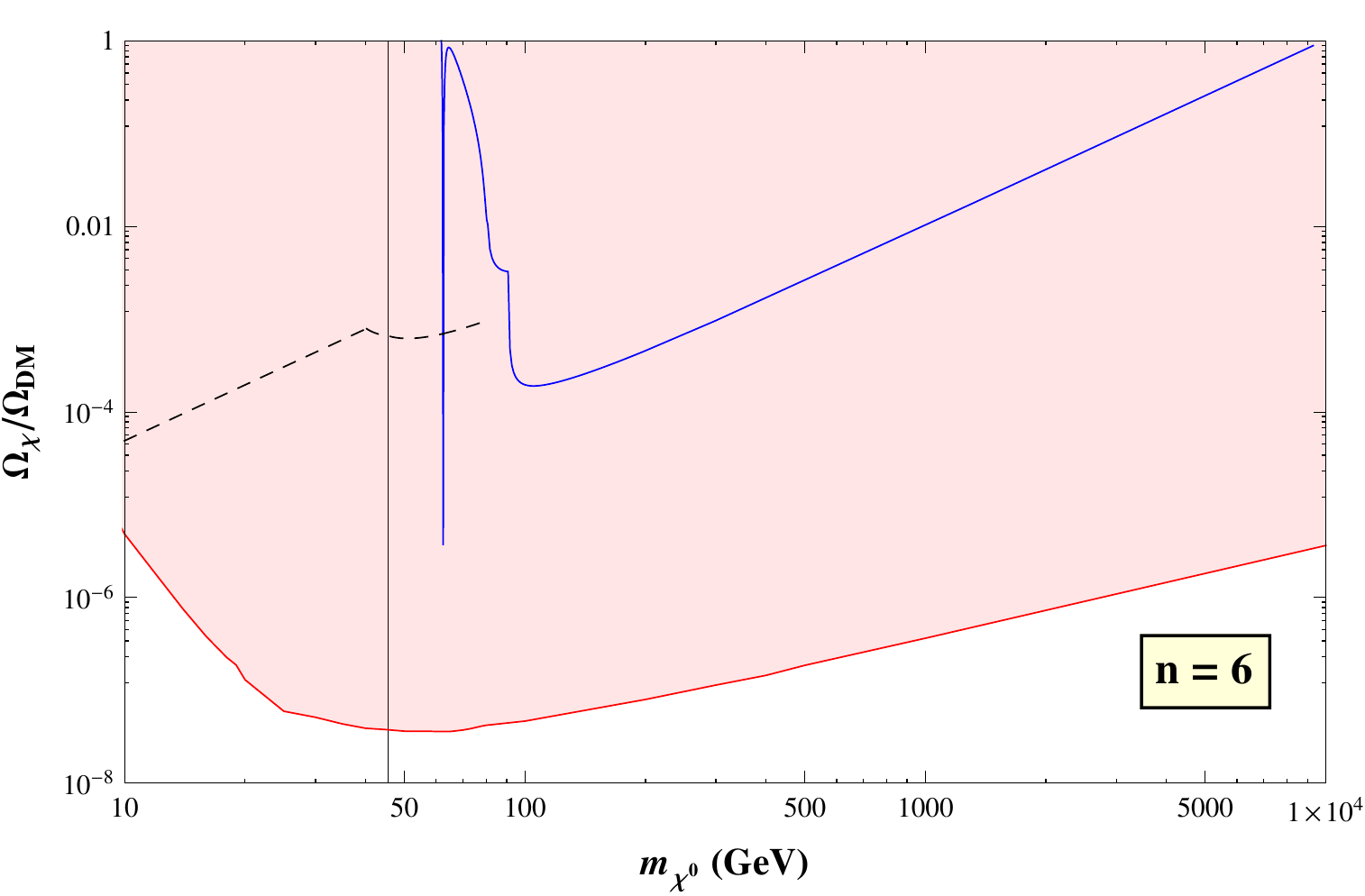}}
\caption{The fraction $\Omega_{\chi}/\Omega_{\rm DM}$ of the total dark matter density for the $n=6$ model as a function of $m_{\chi^0}$.  The shaded area above the red curve is excluded by direct-detection constraints from XENON100 data~\cite{Aprile:2012nq}, conservatively taking $\Lambda_0 = 0$.  The solid blue curve shows the predicted relic density assuming a standard thermal history of the universe, for the parameters $\Lambda_0 = 0.01$, $\lambda_3 = - 0.01$.  The black dashed curve is the relic density in the case that coannihilations are maximal (see text for details).  Masses below $m_Z/2$ (to the left of the vertical black line) are excluded by the LEP constraint on the invisible $Z$ width.}
\label{fig:densityexclusion6}
\end{figure}

\begin{figure}
\resizebox{0.5\textwidth}{!}{\includegraphics{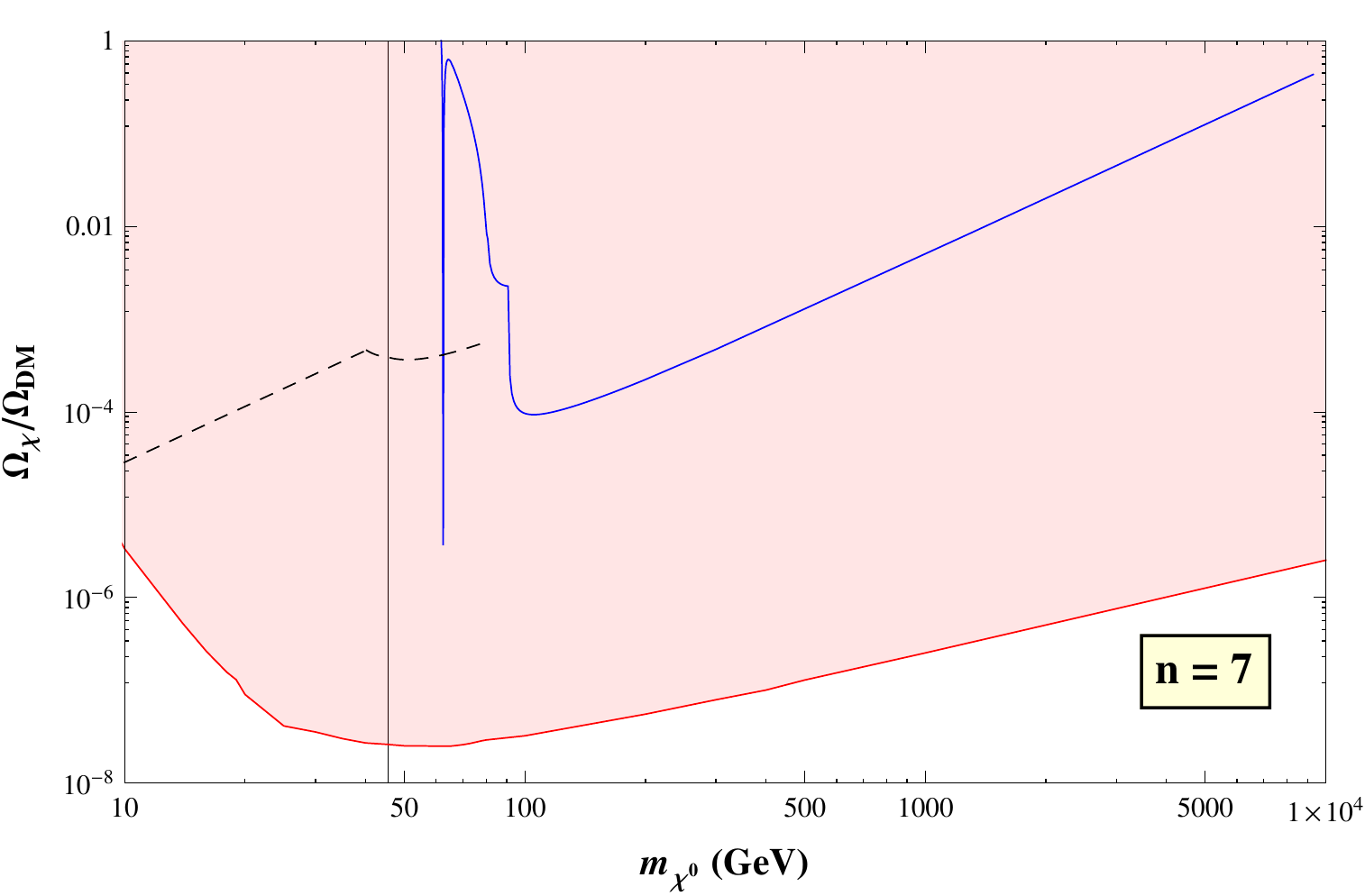}
}\resizebox{0.5\textwidth}{!}{\includegraphics{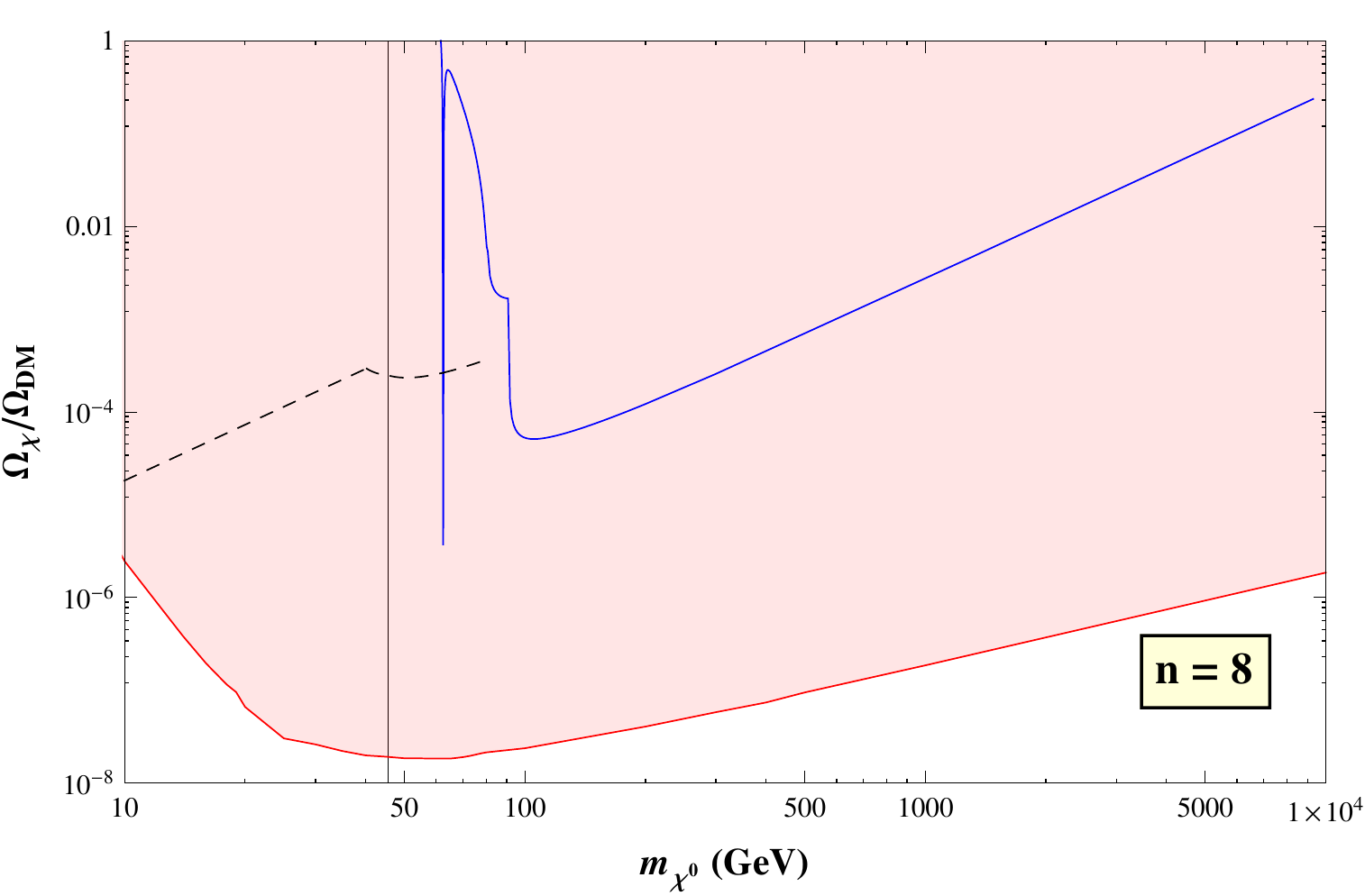}}
\caption{The same as Fig.~\ref{fig:densityexclusion6} but for $n = 7$ (left) and 8 (right).}
\label{fig:densityexclusion78}
\end{figure}

The upper limit on $\Omega_{\chi}/\Omega_{\rm DM}$ from XENON100~\cite{Aprile:2012nq} as a function of the $\chi^0$ mass is shown in Figs.~\ref{fig:densityexclusion6} and~\ref{fig:densityexclusion78} for the multiplets with $n = 6$, 7, and 8 (the limit depends on $n$ like $1/(n-1)^2$).  The shaded area bounded by the red curve is excluded.  Here we have set $\Lambda_0 = 0$ in order to obtain the most conservative limit; taking $\Lambda_0 \neq 0$ has only a tiny effect on the limit.

Finally we comment on the behavior of the upper limit on $\Omega_{\chi}/\Omega_{\rm DM}$ at large $\chi^0$ masses.  The scattering cross section $\sigma^{\chi}_{\rm SI}$ is overwhelmingly dominated by the $Z$-exchange contribution, which is independent of $m_{\chi^0}$ in the large-$m_{\chi^0}$ limit.  The XENON100 collaboration quotes a cross section limit for dark matter masses up to 1000~GeV.  When the dark matter particle mass is large compared to the mass of the target nucleus, the energy transfer for a given target nucleus asymptotes to a constant which depends only on the velocity of the incoming dark matter particle.  The experimental cross section limit then varies inversely with the ambient number density of dark matter particles, which in turn goes like the (fixed) mass density times $1/m_{\chi^0}$.  Therefore, the upper bound on $\Omega_{\chi}/\Omega_{\rm DM}$ grows linearly with $m_{\chi^0}$ for masses that are large compared to the target nucleus mass, and can be extrapolated to arbitrarily heavy masses.

\section{Thermal relic density}
\label{sec:relic}

We now compute the thermal relic density of $\chi^{0} + \chi^{0*}$ and compare it to the upper bound from direct detection found in the previous section.  We assume a standard thermal history of the universe: i.e., that the temperature was high enough at one time for $\chi^{0(*)}$ to have been in thermal equilibrium, that there were no late-decaying relics that would significantly dilute the abundance of $\chi^{0(*)}$, and that there were no late-decaying relics that decayed to states in $X$ and hence boosted the $\chi^{0(*)}$ relic abundance.  

We will show that, for all allowed parameter choices, the thermal relic abundance of $\chi^{0(*)}$ is too large to be consistent with the upper bound on $\Omega_{\chi}/\Omega_{\rm DM}$ from direct detection.  The U(1)-preserving models with $n = 6$, 7, and 8 are thus excluded assuming a standard thermal history.  

The fraction of the dark matter density that is due to $\chi$ is given in terms of the total $\chi^0 \chi^{0*}$ annihilation cross section by
\begin{equation}
	\frac{\Omega_{\chi}}{\Omega_{\rm DM}} = \frac{\langle \sigma v_{rel} \rangle_{\rm std}}
	{\frac{1}{2} \langle \sigma v_{rel} (\chi^0 \chi^{0*} \to \mbox{any})\rangle},
\end{equation}
where $v_{rel}$ is the relative velocity of the two colliding dark matter particles and $\langle \sigma v_{rel}\rangle_{\rm std}$ is the ``standard'' annihilation cross section required to obtain the correct dark matter relic abundance, for which we use $\langle \sigma v_{rel}\rangle_{\rm std} = 3 \times 10^{-26}$~cm$^3/$s~\cite{Steigman:2012nb}.  The brackets indicate an average over the velocity distribution at the time of freeze-out, which is only necessary if the annihilation cross section vanishes in the $v_{rel} \to 0$ limit.  The factor of $1/2$ in the denominator accounts for the probability that, in a collision, any given $\chi$ particle meets one with the opposite U(1) charge so that an annihilation can take place.  This ratio is shown by the solid blue lines in Figs.~\ref{fig:densityexclusion6} and ~\ref{fig:densityexclusion78}.  We explain the ingredients in what follows.

\subsection{Annihilation to two-body final states}

A $\chi^0 \chi^{0*}$ pair can annihilate to the two-body final states $W^+W^-$, $ZZ$, $hh$, and $f \bar f$.  We compute the annihilation cross sections in the zero-velocity limit.  Because these cross sections are all nonzero in this limit, we do not need to average over the velocity distribution.

The annihilation cross sections to two-body final states are given in the $v_{rel} \to 0$ limit by
\begin{eqnarray}
	\sigma v_{rel}(\chi^0 \chi^{0*} \to W^+ W^-) &=& 
	\frac{m_W^4}{8\pi v^4}\sqrt{1 - \frac{m_W^2}{m_{\chi^0}^2}}
	\left[ \frac{A_W^2}{m_{\chi^0}^2} \left(3 - 4\frac{m_{\chi^0}^2}{m_W^2} 
	+ 4\frac{m_{\chi^0}^4}{m_W^4} \right) \right. \nonumber \\
	&& \left. \ + \ 2A_W B_W \left(1 - 3\frac{m_{\chi^0}^2}{m_W^2} 
	+ 2\frac{m_{\chi^0}^4}{m_W^4}\right) 
	+ B_W^2 m_{\chi^0}^2 \left(1 - \frac{m_{\chi^0}^2}{m_W^2}\right)^2 \right], 
	\nonumber \\
	\sigma v_{rel}(\chi^0 \chi^{0*} \to Z Z) &=& 
	\frac{m_Z^4}{16\pi v^4} \sqrt{1 - \frac{m_Z^2}{m_{\chi^0}^2}}
	\left[ \frac{A_Z^2}{m_{\chi^0}^2} \left(3 - 4\frac{m_{\chi^0}^2}{m_Z^2} 
	+ 4\frac{m_{\chi^0}^4}{m_Z^4}\right) \right. \nonumber \\
	&& \left. \ + \ 2A_Z B_Z \left(1 - 3\frac{m_{\chi^0}^2}{m_Z^2} 
	+ 2\frac{m_{\chi^0}^4}{m_Z^4}\right) 
	+ B_Z^2 m_{\chi^0}^2\left(1 - \frac{m_{\chi^0}^2}{m_Z^2}\right)^2 \right],
	\nonumber \\
	\sigma v_{rel}(\chi^0 \chi^{0*} \to hh) &=& \frac{\Lambda_0^2}{64\pi m_{\chi^0}^2}
	\sqrt{1 - \frac{m_h^2}{m_{\chi^0}^2}}
	\left[1 + \frac{3 m_h^2}{4m_{\chi^0}^2 - m_h^2} 
	- \frac{2v^2\Lambda_0}{2m_{\chi^0}^2 - m_h^2} \right]^2,
	\nonumber \\
	\sigma v_{rel}(\chi^0 \chi^{0*} \to f \bar f) &=& \frac{N_c}{4\pi} 
	\left(1 - \frac{m_f^2}{m_{\chi^0}^2}\right)^{3/2}
	\frac{m_f^2 \Lambda_0^2}{(4m_{\chi^0}^2 - m_h^2)^2}.
	\label{eq:annihilation}
\end{eqnarray}
Here the coefficients in the cross sections to $WW$ and $ZZ$ are given by
\begin{eqnarray}
	A_W &=& (n-1) + \frac{v^2 \Lambda_0}{4m_{\chi^0}^2 - m_h^2},
	\qquad \qquad B_W = \frac{4(n-1)}{m_W^2 - m_{\chi^0}^2 - m_{\chi^{+}}^2}, \nonumber \\
	A_Z &=& (n-1)^2 + \frac{v^2\Lambda_0}{4m_{\chi^0}^2 - m_h^2},
	\qquad \qquad B_Z = -\frac{4(n-1)^2}{2m_{\chi^0}^2 - m_Z^2}.
\end{eqnarray}
Diagrams involving $s$-channel $Z$-exchange are zero in the low-energy limit, so that annihilation to $f\bar f$ proceeds only through an $s$-channel Higgs.
We checked our analytic results by implementing the relevant couplings into CalcHEP~\cite{CalcHEP}.

Above threshold, $\chi^{0*} \chi^0 \to ZZ$ has the largest cross section, followed by $\chi^{0*} \chi^0 \to W^+W^-$, which is smaller by about a factor of 20 for $n=6$.  Annihilation rates to $hh$ and $t \bar t$ are much smaller, as can be seen by the fact that their kinematic thresholds are not even visible in Figs.~\ref{fig:densityexclusion6} and~\ref{fig:densityexclusion78}.  The latter two processes are controlled by the coupling $\Lambda_0$; we took the sample value $\Lambda_0 = 0.01$ in Figs.~\ref{fig:densityexclusion6} and~\ref{fig:densityexclusion78}.  As we will see in Sec.~\ref{sec:higgspole}, significantly larger values of $\Lambda_0$ are constrained by the measured rate for $h \to \gamma\gamma$.  $\Lambda_0$ also contributes to the annihilations to $WW$ and $ZZ$ through the $A_W$ and $A_Z$ coefficients; its effect is numerically small and falls with increasing $m_{\chi^0}$.  We also took the sample value $\lambda_3 = -0.01$; $\lambda_3$ has a tiny effect on the annihilation cross section to $WW$ through the $\chi^0$--$\chi^+$ mass splitting.  Overall, the total annihilation cross section above the $WW$ threshold depends very weakly on $\Lambda_0$ and $\lambda_3$, and is instead controlled almost entirely by $n$ and $m_{\chi^0}$.

The cross sections for annihilation to $WW$, $ZZ$, and $hh$ fall like $1/m_{\chi^0}^2$ at large $m_{\chi^0}$, while the cross section to $f \bar f$ falls like $1/m_{\chi^0}^4$.  This leads to the growth of $\Omega_{\chi}/\Omega_{\rm DM}$ proportional to $m_{\chi^0}^2$ for large $m_{\chi^0}$ shown in Figs.~\ref{fig:densityexclusion6} and~\ref{fig:densityexclusion78}.  As we saw in Sec.~\ref{sec:dirdet}, the upper bound on $\Omega_{\chi}/\Omega_{\rm DM}$ from direct detection grows only linearly with $m_{\chi^0}$.  Increasing $m_{\chi^0}$ thus leads only to more severe conflict between the relic abundance and the direct-detection limit.

We finally comment on the possibility that attractive electroweak interactions between $\chi^0$ and $\chi^{0*}$ form bound states that increase the annihilation cross section for $m_{\chi^0} \gg m_W$, an effect known as Sommerfeld enhancement~\cite{Hisano:2006nn}.  This effect was studied in Ref.~\cite{Cirelli:2007xd} for a real scalar 7-plet with hypercharge zero, which found an increase in the mass at which the dark matter candidate obtained the correct relic density by about a factor of 3 (to about 25~TeV), corresponding to about an order of magnitude enhancement of the annihilation cross section during freeze-out.  However, for our models, the direct-detection constraint is $10^5$--$10^6$ times stronger than the perturbative relic density at large $m_{\chi^0} \gtrsim 1$~TeV.  We expect that an enhancement of the annihilation cross section that is sufficiently large to affect our exclusion would be extremely difficult to obtain.

\subsection{Annihilation to off-shell $WW$ below threshold}

Below the $WW$ threshold, the largest two-body annihilation process is $\chi^{0*} \chi^0 \to b \bar b$, which has a small cross section.  Annihilation to off-shell $WW$ can be significantly larger.  We compute the annihilation cross section to off-shell $WW$ by generating $\chi^{0*} \chi^0 \to e^+\nu_e e^- \bar \nu_e$ using CalcHEP~\cite{CalcHEP} and then multiplying by $[1/{\rm BR}(W \to e \nu)]^2 = 81$ at tree level.  We include $b \bar b$ and off-shell $WW$ in the blue solid curves in Figs.~\ref{fig:densityexclusion6} and~\ref{fig:densityexclusion78} for $m_{\chi^0} < m_W$.  We again take $\Lambda_0 = 0.01$ and $\lambda_3 = -0.01$.
Annihilations to off-shell $ZZ$ could also be included; however, their contribution is very small compared to off-shell $WW$ because the $Z$ bosons are further off shell for a given $m_{\chi^0}$.

\subsection{Resonant annihilation through the Higgs pole}
\label{sec:higgspole}

The most interesting feature in the annihilation cross section below the $WW$ threshold is the Higgs pole at $m_{\chi^0} = m_h/2$.  The possibility that this resonant annihilation may suppress the $X$ relic density enough to evade the direct-detection constraints is excluded by a combination of constraints from the oblique parameters and the observed rate for $pp \to h \to \gamma\gamma$ from the LHC~\cite{hgaga-LHC}.  

The $\chi^0 \chi^{0*} \to h \to b \bar b$ annihilation cross section near the Higgs resonance is obtained from the last line of Eq.~(\ref{eq:annihilation}) by making the replacement in the denominator,
\begin{equation}
	(4 m_{\chi^0}^2 - m_h^2)^2 \rightarrow (4 m_{\chi^0}^2 - m_h^2)^2 + m_h^2 \Gamma_h^2.
\end{equation}
Here $\Gamma_h$ is the total decay width of the SM Higgs boson; the contribution from $b \bar b$ final states is given at tree level by~\cite{HHG}
\begin{equation}
	\Gamma(h \to b \bar b) = \frac{N_c m_b^2 m_h}{8 \pi v^2} 
	\left( 1 - \frac{4 m_b^2}{m_h^2} \right)^{3/2},
\end{equation}
where $N_c = 3$ is the number of colors.

\begin{figure}
\resizebox{0.8\textwidth}{!}{\includegraphics{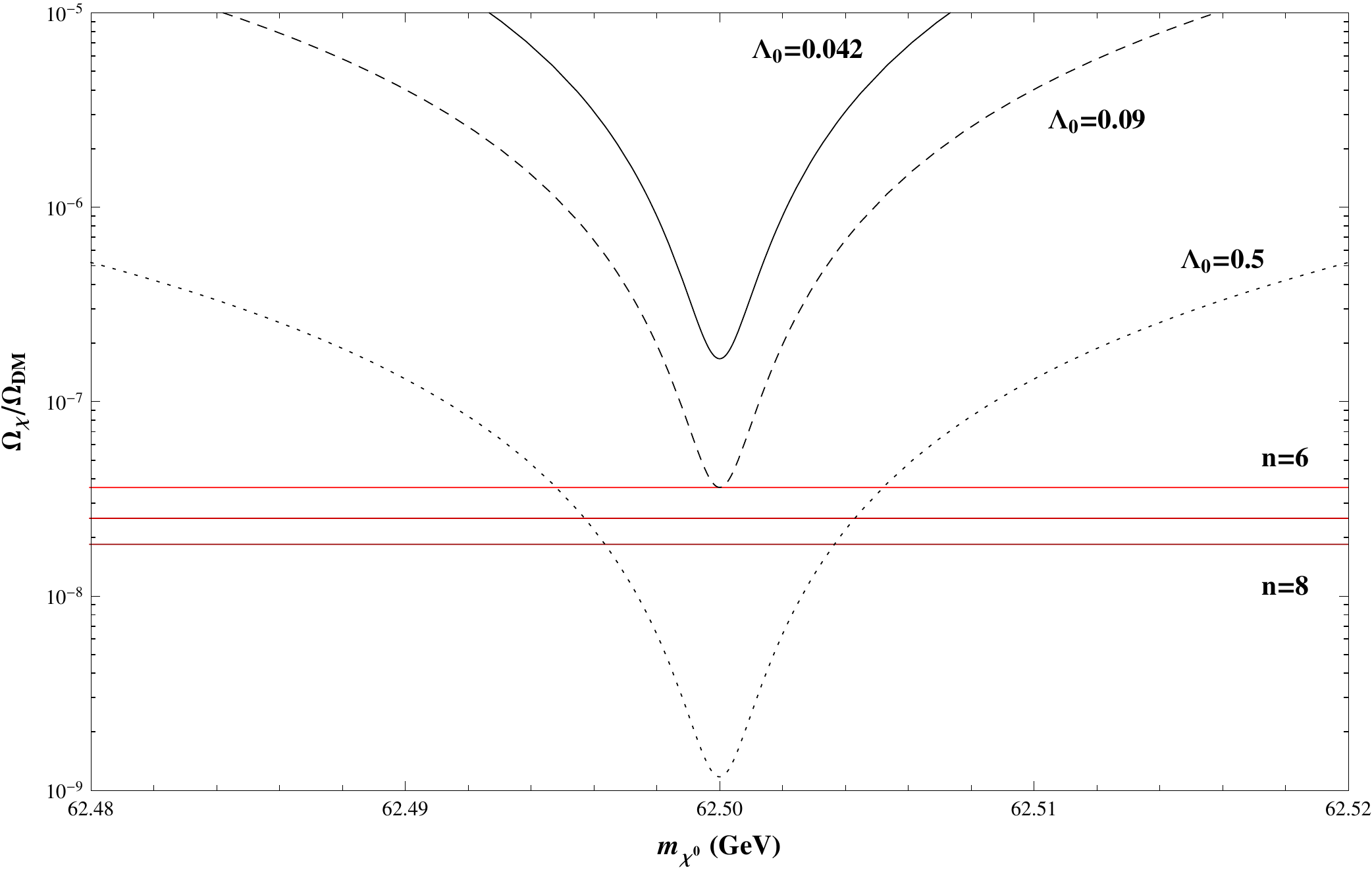}}
\caption{Fractional relic density $\Omega_{\chi}/\Omega_{\rm DM}$ for $m_{\chi^0} \simeq m_h/2$ from $\chi^0 \chi^{0*} \to b \bar b$ annihilation, for various values of $\Lambda_0$ (solid, dashed, and dotted black curves).  Also shown are the upper limits on $\Omega_{\chi}/\Omega_{\rm DM}$ from direct detection for (top to bottom) $n = 6$, 7, and 8 (red horizontal lines).}
\label{fig:bbhiggswidth}
\end{figure}

Note that we can capture the effects of higher-order corrections and additional final states on $\sigma v_{rel}(\chi^0 \chi^{0*} \to h \to {\rm any})$ near the Higgs resonance simply by choosing the value of $m_b$ to yield the correct total SM Higgs width, $\Gamma_h = 4.07 \pm 0.16$~MeV~\cite{Dittmaier:2012vm} for $m_h = 125$~GeV.  We obtain this width from the tree-level formula above when $m_b = 4.08$~GeV.  Using this value of $m_b$, we compute $\Omega_{\chi}/\Omega_{\rm DM}$ in the vicinity of the Higgs resonance using the tree-level $\chi^0 \chi^{0*} \to b \bar b$ cross section and $h \to b \bar b$ partial width.  The result is shown in Fig.~\ref{fig:bbhiggswidth} for various values of $\Lambda_0$.\footnote{When specified in terms of $\Lambda_0$, $\sigma v_{rel} (\chi^0 \chi^{0*} \to b \bar b)$ is independent of $n$.}

\begin{figure}
\resizebox{0.5\textwidth}{!}{\includegraphics{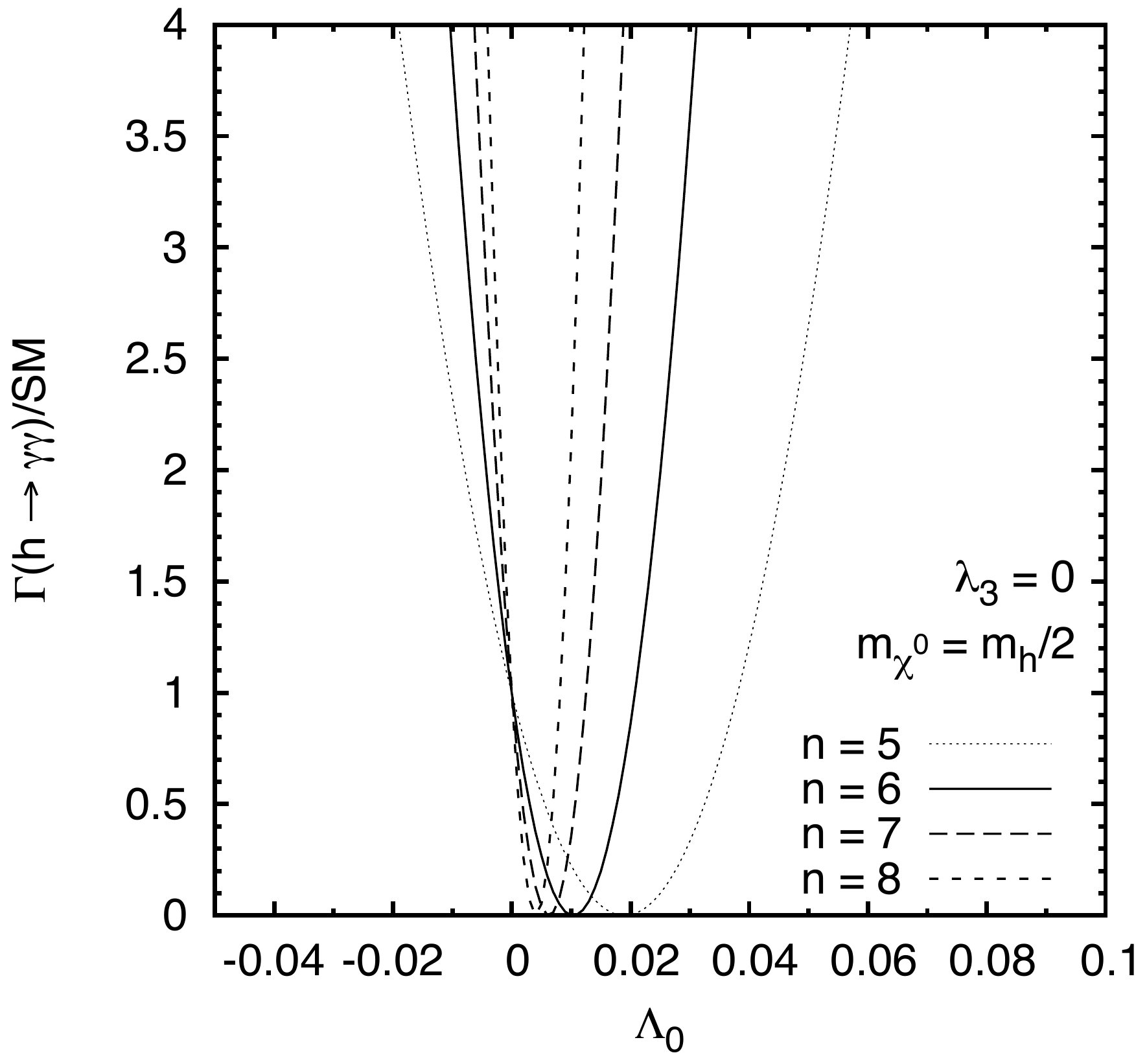}}\resizebox{0.5\textwidth}{!}{\includegraphics{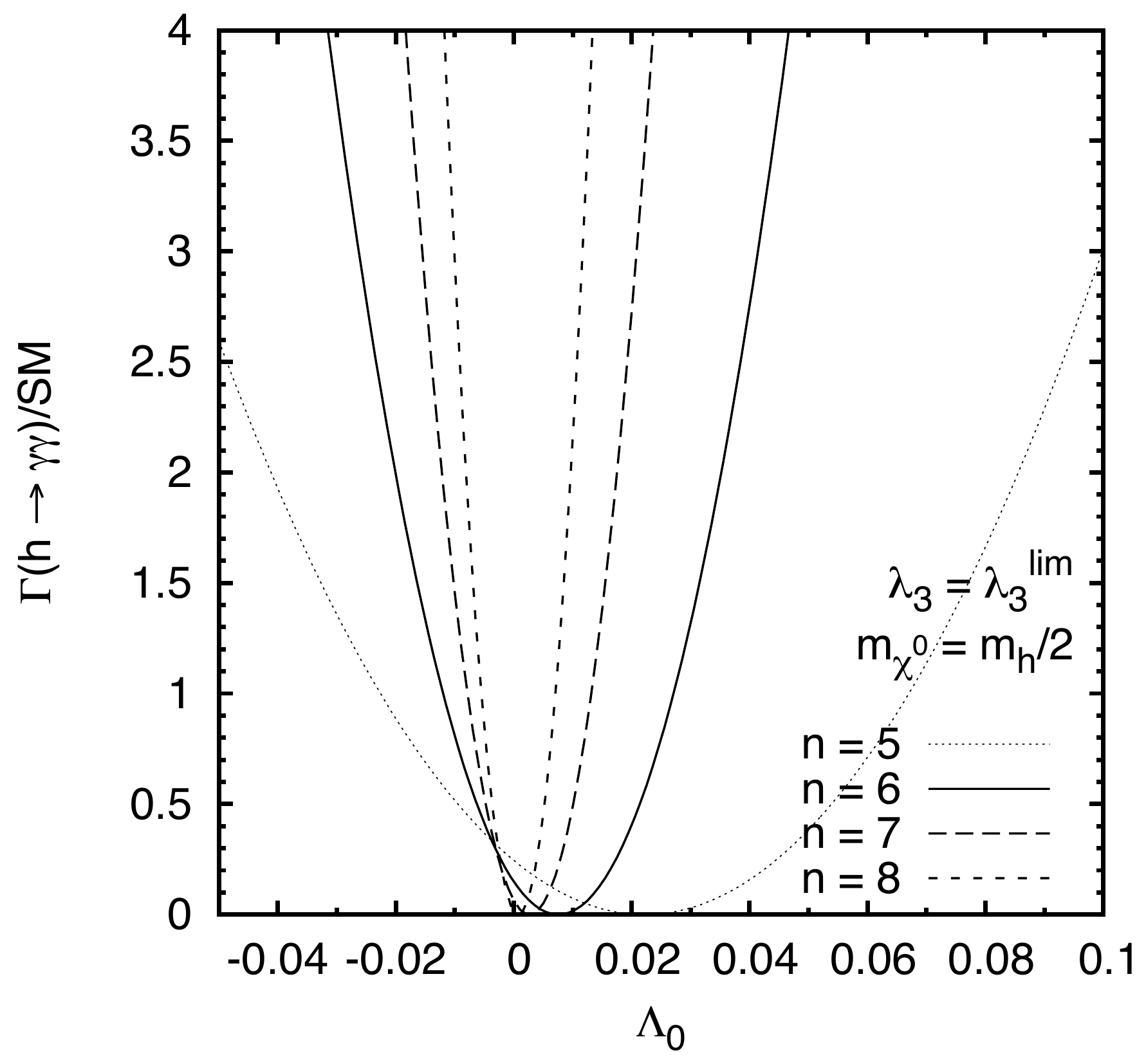}}
\caption{The partial width for $h \to \gamma\gamma$ normalized to its SM value as a function of $\Lambda_0$, for $m_{\chi^0} = m_h/2 = 62.5$~GeV and $\lambda_3 = 0$ (left) and the limiting (minimum) value allowed by the $S$, $T$, and $U$ parameter constraints as given in Table~\ref{tab:lambda3S} (right).}
\label{fig:hgaga}
\end{figure}

The cross section for $\chi^0 \chi^{0*} \to b \bar b$ is proportional to $\Lambda_0^2$, which in turn depends on $\lambda_2$ and $\lambda_3$.  These couplings also control the one-loop contribution of $\chi^Q$ to the Higgs decay to two photons.  Setting $m_{\chi^0} = m_h/2$, the $S$, $T$ and $U$ parameters put an upper bound on $|\lambda_3|$, as summarized in Table~\ref{tab:lambda3S} ($\lambda_3 < 0$ is required in order for $\chi^0$ to be the lightest member of the multiplet).
For $\lambda_3$ within the allowed range, we compute the partial width $\Gamma(h \to \gamma\gamma)$ as a function of $\lambda_2$, leading to a constraint on $\Lambda_0$.  The contributions of the charged $\chi^Q$ states to $\Gamma(h \to \gamma\gamma)$ are summarized in Appendix~\ref{sec:hgaga}.  This partial width normalized to its SM value is shown as a function of $\Lambda_0$ in Fig.~\ref{fig:hgaga} for the two extreme cases, $\lambda_3 = 0$ and the limiting value allowed by the $S$, $T$, and $U$ parameter constraints.  

\begin{table}
\begin{tabular}{cc}
\hline\hline
$n$ & $\lambda_3^{\rm lim}$ \\
\hline
\ 5 \ &  $-1.4 \times 10^{-2}$\\
6 &  $-6.0 \times 10^{-3}$\\
7 &  $-3.3 \times 10^{-3}$\\
8 &  $-2.0 \times 10^{-3}$\\
\hline\hline
\end{tabular}
\caption{The 95\% confidence level lower limit on $\lambda_3$ from the $S$, $T$, and $U$ parameter constraints as a function of the size of the multiplet, for $m_{\chi^0} = m_h/2 = 62.5$~GeV.}
\label{tab:lambda3S}
\end{table}

We see that increasing $|\Lambda_0|$ quickly drives $\Gamma(h \to \gamma\gamma)$ to unacceptably large values.  In particular, $|\Lambda_0| \gtrsim 0.09$ -- which is required to evade the direct-detection limit for the $n = 6$ model -- is ruled out by the LHC measurement of the signal strength for $pp \to h \to \gamma\gamma$.\footnote{In the models considered here, the Higgs couplings are identical to those in the SM except for the contributions to the loop-induced $h \gamma\gamma$ and $h \gamma Z$ couplings from loops of charged scalars.  Because the decays to $\gamma\gamma$ and $\gamma Z$ contribute only a tiny fraction of the Higgs total width, the LHC Higgs signal strength in the $\gamma\gamma$ channel is given to a very good approximation by $\mu \equiv \sigma(pp \to h \to \gamma\gamma)/\sigma_{SM}(pp \to h \to \gamma\gamma) \simeq \Gamma(h \to \gamma\gamma)/\Gamma_{SM}(h \to \gamma\gamma)$.}
Allowing a very generous factor of three enhancement over the SM prediction (already excluded at more than 2$\sigma$ by both ATLAS and CMS~\cite{hgaga-LHC}) requires $|\Lambda_0| < 0.042$ for $n=6$; the limit becomes more stringent for larger $n$.  After applying these constraints, we see that resonant annihilation through the Higgs pole cannot suppress $\Omega_{\chi}/\Omega_{\rm DM}$ enough to allow us to evade the direct-detection limits.

\subsection{Coannihilations}

We finally consider the possibility that the mass splittings among the states $\chi^Q$ are very small,  as is favored by the oblique parameter constraints for low $\chi^0$ masses.  In this case, all members of the multiplet can be present in the thermal bath during freeze-out.  This opens the possibility of annihilations involving the electrically charged states $\chi^Q$ to $\gamma\gamma$, $Z\gamma$, and $W^{\pm} \gamma$ final states,\footnote{The cross section for $\chi^Q \chi^{Q*} \to h\gamma$ vanishes in the zero-velocity limit.} which are on shell below the $WW$ threshold.  Such electroweak-strength annihilation cross sections to two-body final states can easily dominate over those to three-body final states and the bottom-Yukawa-suppressed annihilation cross section to $b \bar b$.

We evaluate the potential impact of such coannihilations by considering the extreme case in which all states $\chi^{Q(*)}$ are present with equal abundances during freeze-out.  Two conditions are required: the mass splittings must be very small so that the equilibrium thermal abundances of each species are equal, and the quartic couplings multiplying operators $\mathcal{O}(X^4)$ must be large enough to maintain equal abundances of states with each charge, even though the annihilation cross section to SM states is different for states with different charge.  For larger mass splittings, coannihilations become less important, and the situation relaxes to our original analysis.

In the limit of equal abundances of all states $\chi^{Q(*)}$, the fraction of the dark matter density due to $\chi$ is given in terms of the annihilation cross sections by
\begin{equation}
	\frac{\Omega_\chi}{\Omega_{\rm DM}} = \frac{\langle \sigma v_{rel} \rangle_{\rm std}}
	{\frac{1}{2} \frac{1}{n^2} \sum_{Q_1, Q_2 \geq 0} 
	\langle \sigma v_{rel}(\chi^{Q_1} \chi^{Q_2 *} \to \mbox{any})\rangle}.
	\label{eq:coanndensity}
\end{equation}
Note in particular the new factor $1/n^2$ in the denominator, which represents the average over initial charge states for a multiplet of size $n$.  The sum runs over all initial charge combinations; only the combinations with $Q_1 - Q_2 = 0$ and $\pm 1$ contribute below the $WW$ threshold.  We include annihilations to the two-body final states $\gamma\gamma$, $Z\gamma$, $W^{\pm}\gamma$, and $b \bar b$.  Cross section formulas are given in Appendix~\ref{sec:coann}.  The cross sections for the first three processes are independent of $\Lambda_0$ and depend on $\lambda_3$ only through the masses $m_{\chi^Q}$.  For the scalar couplings we set $\Lambda_0 = 0.01$ as usual but we now take $\lambda_3 = 0$, corresponding to degenerate masses for all $\chi^Q$.

The resulting relic abundance of $\chi$ in the full coannihilation case is shown by the black dashed lines in Figs.~\ref{fig:densityexclusion6} and~\ref{fig:densityexclusion78}.  The relic abundance is dramatically reduced compared to that obtained considering only $\chi^0 \chi^{0*}$ annihilations.  However, coannihilations still do not allow us to evade the direct-detection limits: the relic density of $\chi$ remains about four orders of magnitude above the direct-detection exclusion bound for $m_Z/2 \leq m_{\chi^0} \leq m_W$.

\section{Conclusions}
\label{sec:conclusions}

In this paper we studied the class of models in which the SM Higgs sector is extended by a single large complex scalar multiplet in such a way that the Higgs potential preserves an accidental\footnote{We do not consider models where the additional U(1) symmetry is imposed by hand, such as the $Y=0$ case outlined in Ref.~\cite{Joachim}.} global U(1) symmetry at the renormalizable level. The accidental U(1) symmetry is present when $n = 2T + 1 \geq 5$.  Perturbative unitarity of weak-interaction scattering amplitudes involving the large multiplet excludes multiplets with $n > 8$.  We choose the hypercharge to be $Y = 2T$ so that the lightest member of the large multiplet can be electrically neutral. 

We showed that the models with $n = 6$, 7, or 8 are excluded by the incompatibility of the standard thermal freeze-out relic density with the dark matter direct-detection cross section limit, assuming a standard thermal history of the universe.  The model with $n = 5$ evades the direct-detection constraint because its lightest state can decay via a Planck-suppressed dimension-5 operator during the first few days to few years after the big bang.

This exclusion can be evaded if the model is modified in such a way as to break the global U(1) symmetry.  One approach is to add one or more additional, smaller scalar multiplets in such a way that the accidental global symmetry is eliminated.  This induces effective higher-dimensional operators involving the SM Higgs field and the large multiplet $X$ that break the global U(1).  Such operators typically also induce a vev for $X$ and mixing between the states of $X$ and the SM Higgs doublet.  Models of this type involving a large multiplet with $n=7$ and $Y=4$ (whose vev preserves $\rho = 1$ at tree level~\cite{rhomults,HHG}) have recently been discussed in Ref.~\cite{Hisano:2013sn}.

A second approach is to arrange the model in such a way that the global U(1) is broken down to $Z_2$.  In this case, the real and imaginary components of the neutral member of $X$ can be arranged to have different masses, so that the $Z$-mediated direct-detection scattering process $\chi^{0,r} N \to \chi^{0,i} N$ is kinematically forbidden.  The $h$-mediated process $\chi^{0,r} N \to \chi^{0,r} N$ has a much smaller cross section and remains experimentally viable.  Such a model can be constructed for a large multiplet with $n = 6$ or $n = 8$ if its hypercharge is chosen as $Y = 1$.  Then the U(1)-breaking operator $(\Phi^{\dagger} \tau^a \widetilde \Phi)(\widetilde X^{\dagger} T^a X)$ appears in the scalar potential, where $\widetilde \Phi$ and $\widetilde X$ denote the conjugate multiplets.  We will address such models in Ref.~\cite{Z2}.

\begin{acknowledgments}
We thank Thomas Gr\'egoire for helpful discussions on the unitarity constraints.
This work was supported by the Natural Sciences and Engineering Research Council of
Canada.
\end{acknowledgments}

\appendix
\section{Feynman rules}

We summarize the Feynman rules for the couplings of $\chi$ states to gauge and Higgs bosons.  The following are for all particles and momenta incoming.  We take $Q \geq 0$ and denote the antiparticle of $\chi^Q$ as $\chi^{Q*}$.

Couplings to one or two Higgs bosons are as follows:
\begin{eqnarray}
	\chi^{Q*} \chi^{Q} h &=& -i v
		\left[\lambda_2 - \frac{1}{2}\lambda_3\left(Q - \frac{n-1}{2}\right)\right] 
		= -i v \Lambda_Q \nonumber \\
	\chi^{Q*} \chi^{Q} hh &=& -i \left[\lambda_2 
		- \frac{1}{2}\lambda_3\left(Q - \frac{n-1}{2}\right)\right]
		= -i \Lambda_Q.
		\label{eq:xxh}
\end{eqnarray}
Couplings to two gauge bosons are as follows:
\begin{eqnarray}
	\chi^{Q*} \chi^{Q} W^+_\mu W^-_\nu &=& \frac{i e^2}{s_W^2}
		\left[(2Q+1)\frac{(n-1)}{2}-Q^2\right] g_{\mu\nu} \nonumber \\
	\chi^{(Q+2)*} \chi^{Q} W^+_\mu W^+_\nu &=& \frac{i e^2}{s_W^2}
		\sqrt{(Q+2)(n-2-Q)(Q+1)(n-1-Q)} \, g_{\mu\nu} 
		= \chi^{Q*} \chi^{Q+2} W^-_\mu W^-_\nu \nonumber \\
	\chi^{Q*} \chi^{Q} A_\mu A_\nu &=& 2 i e^2 Q^2 g_{\mu\nu} \nonumber \\
	\chi^{Q*} \chi^{Q} Z_\mu Z_\nu &=& \frac{2 i e^2}{s_W^2 c_W^2}
		\left[c_W^2 Q - \frac{n - 1}{2}\right]^2 g_{\mu\nu} \nonumber \\
	\chi^{Q*} \chi^{Q+1} W^-_\mu A_\nu &=& \frac{i e^2}{s_W}
		\sqrt{\frac{(Q+1)(n-1-Q)}{2}}(1+2Q) g_{\mu\nu} 
		= \chi^{(Q+1)*} \chi^{Q} W^+_\mu A_\nu \nonumber \\
	\chi^{Q*} \chi^{Q+1} W^-_\mu Z_\nu &=& \frac{i e^2}{s_W^2 c_W}
		\sqrt{\frac{(Q+1)(n-1-Q)}{2}} \left[(1+2Q)c_W^2 - n + 1\right] g_{\mu\nu} 
		= \chi^{(Q+1)*} \chi^{Q} W^+_\mu Z_\nu  \nonumber \\
	\chi^{Q*} \chi^{Q} Z_\mu A_\nu &=& \frac{2i e^2 Q}{s_W c_W}
		\left[c_W^2 Q - \frac{n - 1}{2}\right] g_{\mu\nu}.
\end{eqnarray}
Couplings to one gauge boson are as follows:
\begin{eqnarray}
	\chi^{Q*}(p) \chi^{Q+1}(k) W_\mu^- &=& -\frac{i e}{s_W} \sqrt{\frac{(Q+1)(n-1-Q)}{2}}
		\left(k - p\right)_\mu 
		= \chi^{(Q+1)*}(p) \chi^{Q}(k) W_\mu^+ \nonumber \\
	\chi^{Q*}(p) \chi^{Q}(k) A_\mu &=& -i e Q \left(k-p\right)_\mu \nonumber \\
	\chi^{Q*}(p) \chi^{Q}(k) Z_\mu &=& -\frac{i e}{s_W c_W} 
		\left[c_W^2 Q - \frac{n - 1}{2}\right] \left(k - p\right)_\mu.
\end{eqnarray}

\section{Contributions to the oblique parameters from a scalar electroweak multiplet}
\label{sec:oblique}

For a multiplet of hypercharge $Y$ and size $n = 2T+1$, the contribution to the $S$ parameter is given by~\cite{Lavoura:1993nq}
\begin{equation}
	S = \frac{4 s_W^2 c_W^2}{\alpha} 
	\left[\Pi^{\prime}_{ZZ}(0) 
	- \frac{c_W^2 - s_W^2}{c_W s_W}\Pi^{\prime}_{Z\gamma}(0)
	- \Pi^{\prime}_{\gamma\gamma}(0)\right] 
	= - \frac{Y}{6\pi}\sum_{i=0}^{n-1}T_i^3 \log (m_i^2),
\end{equation}
where $m_i$ and $T^3_i$ denote the mass and third component of isospin of the complex scalar mass eigenstate $\chi_i$.\footnote{Note that we use the convention $Q=T^3+Y/2$ and as such $Y$ in the results of~\cite{Lavoura:1993nq} must be replaced by $Y/2$.}

The contribution to the $T$ parameter can be similarly represented as~\cite{Lavoura:1993nq}
\begin{eqnarray}
	T &=& \frac{1}{\alpha} \left[\frac{\Pi_{WW}(0)}{m_W^2}
	- \frac{\Pi_{ZZ}(0)}{m_Z^2}\right] \nonumber \\
	&=& \frac{1}{4\pi m_Z^2 s_W^2 c_W^2}
	\left[\sum_{i=0}^{n-1} m_i^2 \log (m_i^2) \left[T(T+1)-(T^3_i)^2\right] \right. \nonumber \\ 
	&& \left. \qquad
	- \sum_{i=0}^{n-2} (T-T_i^3) (T+T_i^3+1) f_2(m_i,m_{i+1}) \right],
\end{eqnarray}
where
\begin{equation}
	f_2(m_1,m_2) = \int_0^1 dx\,\left[x m_1^2+(1-x)m_2^2\right]
	\log\left[x m_1^2+(1-x)m_2^2\right].
\end{equation}

Finally, the contribution to the $U$ parameter is~\cite{Lavoura:1993nq}
\begin{eqnarray}
	U &=& \frac{4 s_W^2}{\alpha} \left[\Pi^{\prime}_{WW}(0)
	- c_W^2 \Pi^{\prime}_{ZZ}(0) 
	- 2 s_W c_W \Pi^{\prime}_{Z\gamma}(0)
	- s_W^2 \Pi^{\prime}_{\gamma\gamma}(0)\right] \nonumber \\
	&=& \frac{1}{\pi} \left[ \sum_{i=0}^{n-2}(T-T_i^3)(T+T_i^3+1)\,f_1(m_i,m_{i+1})-\frac{1}{3}\sum_{i=0}^{n-1} (T^3_i)^2 \log (m_i^2)\right],
\end{eqnarray}
where
\begin{equation}
	f_1(m_1,m_2) = \int_0^1 dx\,x(1-x) \log\left[x\,m_1^2+(1-x)m_2^2\right].
\end{equation}

We constrain these contributions to oblique parameters using a $\chi^2$ variable including the correlations in the measured $S$, $T$, and $U$ values,
\begin{equation}
	\chi^2 = \sum_{i,j} (\mathcal{O}_i - \mathcal{O}_i^{\rm exp})
	(\mathcal{O}_j - \mathcal{O}_j^{\rm exp}) [\sigma^2]^{-1}_{ij},
\end{equation}
where $\mathcal{O}_i$ is the $i$th observable and $[\sigma^{2}]^{-1}_{ij}$ is the inverse of the matrix of uncertainties,
\begin{equation}
	[\sigma^2]_{ij} = \Delta \mathcal{O}_i \, \Delta \mathcal{O}_j \, \rho_{ij},
\end{equation}
where $\rho_{ij}$ are the relative correlations (note $\rho_{ii} = 1$).  For the three-observable case of interest, we can invert the matrix $\sigma^2$ explicitly and write
\begin{eqnarray}
	\chi^2 &=& \frac{1}{\left(1 - \rho_{ST}^2 - \rho_{TU}^2 - \rho_{SU}^2 
	+ 2\rho_{ST}\rho_{TU}\rho_{SU}\right)}
	\left[ \frac{\left(1 - \rho_{TU}^2 \right)\left(S-S_{\rm exp}\right)^2}
	{\left( \Delta S_{\rm exp} \right)^2}
	+ \frac{\left(1 - \rho_{SU}^2 \right)\left(T-T_{\rm exp}\right)^2}
	{\left( \Delta T_{\rm exp} \right)^2}
	\right.\nonumber\\
	&&\left. + \frac{\left(1 - \rho_{ST}^2 \right)\left(U-U_{\rm exp}\right)^2}
	{\left( \Delta U_{\rm exp} \right)^2}
	- 2 \left(\rho_{ST}-\rho_{TU}\rho_{SU}\right)
	\frac{\left(S-S_{\rm exp}\right)\left(T-T_{\rm exp}\right)}
	{\Delta S_{\rm exp}\,\Delta T_{\rm exp}}
	\right.\nonumber\\
	&&\left. - 2 \left(\rho_{TU}-\rho_{ST}\rho_{SU}\right)
	\frac{\left(T-T_{\rm exp}\right)\left(U-U_{\rm exp}\right)}
	{\Delta T_{\rm exp}\,\Delta U_{\rm exp}}
	- 2 \left(\rho_{SU}-\rho_{TU}\rho_{ST}\right)
	\frac{\left(S-S_{\rm exp}\right)\left(U-U_{\rm exp}\right)}
	{\Delta S_{\rm exp}\,\Delta U_{\rm exp}}\right].
\end{eqnarray}
Here $S_{\rm exp}$, $T_{\rm exp}$, and $U_{\rm exp}$ are the experimental central values, $\Delta S_{\rm exp}$, $\Delta T_{\rm exp}$ and $\Delta U_{\rm exp}$ are their $1\sigma$ experimental uncertainties, $\rho_{ST}$, $\rho_{SU}$, and $\rho_{TU}$ are their relative correlations, and $S$, $T$, and $U$ are the contributions from the scalar multiplet computed using the formulas above.

\section{Contribution to $h \to \gamma\gamma$}
\label{sec:hgaga}

The experimental observation of $pp \to h \to \gamma\gamma$ with a rate close to its
SM value allows us to constrain the strength of the $\chi^0 \chi^{0*} h$ coupling in the Higgs pole annihilation region, $m_{\chi^0} \sim m_h/2$.  The charged members of the $X$ multiplet contribute to the loop-induced $h \to \gamma\gamma$ partial width~\cite{HHG},
\begin{equation}
	\Gamma(h \to \gamma\gamma) = \frac{\alpha^2 g^2}{1024 \pi^3} \frac{m_h^3}{m_W^2}
	\left| \sum_i N_{ci} Q_i^2 F_i(\tau) \right|^2,
\end{equation}
where $i$ runs over charged particles of spin $1$, $1/2$, and $0$, $Q$ is the electric charge in units of $e$, $N_{ci}$ is the color multiplicity and the functions $F_i(\tau)$ depend on the particle's spin,
\begin{eqnarray}
	F_1&=&2+3\tau+3\tau(2-\tau) f(\tau)\nonumber \\
	F_{1/2}&=&-2\tau[1+(1-\tau) f(\tau)]\nonumber \\
	F_0&=&\beta\tau[1-\tau f(\tau)].
\end{eqnarray}
Here $\tau=4 m_i^2/m^2_h$, and the function $f(\tau)$ is given by
\begin{equation}
f(\tau)=
\left\{
\begin{array}{c c}
\left[\arcsin\left(\sqrt{\frac{1}{\tau}}\right)\right]^2 &\quad  {\rm if}\,\,\tau\geq 1\\
-\frac{1}{4}\left[\ln\left(\frac{\eta_+}{\eta_-}-i\,\pi\right)\right]^2 &\quad  {\rm if}\,\,\tau< 1, \\
\end{array}
\right.
\end{equation}
where we have defined $\eta_\pm = 1\pm\sqrt{1-\tau}$.

For the scalars, the coupling to the Higgs is parameterized by
\begin{equation}
	\beta_i = \frac{m^2_i \ {\rm due \ to \ Higgs}}{m^2_i} 
	= \frac{v^2(\lambda_2/2 - \lambda_3 T^3_i/4)}
		{M^2 + v^2(\lambda_2/2 - \lambda_3 T^3_i/4)}
	= \frac{v^2 \Lambda_Q/2}{M^2 + v^2 \Lambda_Q/2},
\end{equation}
where $T^3_i$ is the third component of isospin of the scalar $\chi_i$ and $\Lambda_Q$ 
is defined in Eq.~(\ref{eq:xxh}).

\section{Cross sections for coannihilations}
\label{sec:coann}

The cross sections for annihilations of charged $\chi$ states into $\gamma\gamma$, $Z\gamma$, $W^{\pm} \gamma$, and $f \bar f$ relevant for coannihilations below the $WW$ threshold are given by
\begin{eqnarray}
	\sigma v_{rel}(\chi^{Q*} \chi^Q \to \gamma\gamma) 
	&=& \frac{e^4 Q^4}{8 \pi m_{\chi^Q}^2} \nonumber \\
	\sigma v_{rel}(\chi^{Q*} \chi^Q \to Z \gamma)
	&=& \frac{e^4 Q^2}{4 \pi m_{\chi^Q}^2 s_W^2 c_W^2} 
	\left[ Q c_W^2 - \frac{(n-1)}{2} \right]^2 \left( 1 - \frac{m_Z^2}{4 m_{\chi^Q}^2} \right)
	\nonumber \\
	\sigma v_{rel}(\chi^{Q*} \chi^{Q+1} \to W^+ \gamma) 
	&=& \frac{e^4 (Q+1) (n - 1 - Q) (2Q + 1)^2}{32 \pi m_{\chi^Q} m_{\chi^{Q+1}} s_W^2}
	\left( 1 - \frac{m_W^2}{(m_{\chi^Q} + m_{\chi^{Q+1}})^2} \right) \nonumber \\
	&=& \sigma v_{rel}(\chi^{(Q+1)*} \chi^Q \to W^- \gamma) \nonumber \\
	\sigma v_{rel}(\chi^{Q*} \chi^Q \to f \bar f) 
	&=& \frac{N_c}{4 \pi} \left( 1 - \frac{m_f^2}{m_{\chi^Q}^2} \right)^{3/2}
	\frac{m_f^2 \Lambda_Q^2}{(4 m_{\chi^Q}^2 - m_h^2)^2},
\end{eqnarray}
where $\Lambda_Q$ was defined in Eq.~(\ref{eq:xxh}).


\end{document}